\documentclass[aps,floatfix,prd,showpacs,superscriptaddress,preprintnumbers]{revtex4}
\usepackage{graphicx}
\usepackage{enumerate}
\usepackage{dcolumn}
\usepackage{bm}

\begin{document}

\title{Issues and Opportunities in Exotic Hadrons\footnote{Corresponding authors: J.J. Dudek (dudek@jlab.org), R.E. Mitchell (remitche@indiana.edu), E.S. Swanson (swansone@pitt.edu).}}

\preprint{INT-PUB-15-066}
 
\author{R.A.  Brice\~{n}o}
\affiliation{
Thomas Jefferson National Accelerator Facility,
12000 Jefferson Avenue, Newport News, VA 23606, USA
}
\affiliation{
Department of Physics, Old Dominion University, Norfolk, VA 23529, USA
}
\author{T.D. Cohen}
\affiliation{
Maryland Center for Fundamental Physics, University of Maryland, College Park, MD, USA}
\author{S. Coito}
\affiliation{
Institute of Modern Physics, CAS, Lanzhou 730000, China
}
\author{J.J. Dudek}
\affiliation{
Thomas Jefferson National Accelerator Facility,
12000 Jefferson Avenue, Newport News, VA 23606, USA
}
\affiliation{
Department of Physics, Old Dominion University, Norfolk, VA 23529, USA
}
\author{E. Eichten}
\affiliation{
Theoretical Physics Department, Fermilab, IL 60510, USA
}
\author{C.S. Fischer}
\affiliation{
Institut f\"{u}r Theoretische Physik, Justus-Liebig-Universit\"{a}t Gie{\ss}en, Heinrich-Buff-Ring 16, D-35392 Gie{\ss}en, Germany.
}
\author{M. Fritsch}
\affiliation{
Helmholtz Institute Mainz, Johann-Joachim-Becher-Weg 45, D-55099 Mainz, Germany}
\affiliation{
Johannes Gutenberg University of Mainz, Johann-Joachim-Becher-Weg 45, D-55099 Mainz, Germany
}
\author{W. Gradl}
\affiliation{
Johannes Gutenberg University of Mainz, Johann-Joachim-Becher-Weg 45, D-55099 Mainz, Germany
}
\author{A. Jackura}
\affiliation{
Physics Department, Indiana University, Bloomington, IN 47405, USA
}
\author{M. Kornicer}
\affiliation{
University of Hawaii, Honolulu, Hawaii 96822, USA
}
\author{G. Krein}
\affiliation{
Instituto de F\'{i}sica Te\'{o}rica, Universidade Estadual Paulista,
Rua Dr. Bento Teobaldo Ferraz, 271 - Bloco II, 01140-070 S\~{a}o Paulo, SP, Brazil
}
\author{R.F. Lebed}
\affiliation{
Department of Physics, Arizona State University, Tempe, Arizona 85287-1504, USA
}
\author{F.A. Machado}
\affiliation{
Department of Physics and Astronomy,
University of Pittsburgh,
Pittsburgh, PA 15260,
USA.
}
\author{R.E. Mitchell}
\affiliation{
Indiana University, Bloomington, Indiana 47405, USA
}
\author{C.J. Morningstar}
\affiliation{
Department of Physics, Carnegie Mellon University, Pittsburgh, PA 15213, USA
}
\author{M. Peardon}
\affiliation{
School of Mathematics, Trinity College, Dublin 2, Ireland
}
\author{M.R. Pennington}
\affiliation{
Thomas Jefferson National Accelerator Facility,
12000 Jefferson Avenue, Newport News, VA 23606, USA
}
\author{K. Peters}
\affiliation{
GSI Helmholtzcentre for Heavy Ion Research GmbH, D-64291 Darmstadt, Germany
}
\author{J.-M. Richard}
\affiliation{
Universit\'{e} de Lyon, Institut de Physique Nucl\'{e}aire de Lyon, IN2P3-CNRS-UCBL, 4, rue Enrico Fermi, Villeurbanne, France
}
\author{C.-P. Shen}
\affiliation{
Beihang University, Beijing 100191, People’s Republic of China
}
\author{M.R. Shepherd}
\affiliation{
Indiana University, Bloomington, Indiana 47405, USA
}
\author{T. Skwarnicki}
\affiliation{
Syracuse University, Syracuse, NY, USA
}
\author{E.S. Swanson}
\affiliation{
Department of Physics and Astronomy,
University of Pittsburgh,
Pittsburgh, PA 15260,
USA.
}
\author{A.P. Szczepaniak}
\affiliation{
Theory Center, Thomas Jefferson National Accelerator Facility,
12000 Jefferson Avenue, Newport News, VA 23606, USA
}
\affiliation{
Center for Exploration of Energy and Matter, Indiana University, Bloomington, IN 47403, USA
}
\affiliation{
Physics Department, Indiana University, Bloomington, IN 47405, USA
}
\author{C.-Z. Yuan}
\affiliation{
Institute of High Energy Physics, Beijing 100049, People’s Republic of China
}

\date{\today}

\begin{abstract}
The last few years have been witness to a proliferation of new results concerning heavy exotic hadrons.  Experimentally, many new signals have been discovered that could be pointing towards the existence of tetraquarks, pentaquarks, and other exotic configurations of quarks and gluons.  Theoretically, advances in lattice field theory techniques place us at the cusp of understanding complex coupled-channel phenomena, modelling grows more sophisticated, and effective field theories are being applied to an ever greater range of situations. It is thus an opportune time to evaluate the status of the field.  In the following, a series of high priority experimental and theoretical issues concerning heavy exotic hadrons is presented.  
\end{abstract}

\maketitle

\section{Introduction}

In 2007 the Belle Collaboration claimed the discovery of the Z(4430). This state attracted considerable attention because it is charged and couples to charmonium, implying that the most economical interpretation of its quark content is $c\bar c \bar{u} d$. The recent high statistics confirmation of the $Z$ by the LHCb collaboration, and the startling demonstration of phase motion, has brought sharp focus on exotic hadronic matter. Indeed, the $Z(4430)$ joins a long list of other putative exotic states, several of which have been reported within the past year:

\begin{quote}
\begin{description}
\item[$c\bar c$ multiquarks] \hfill \\
X(3872), Z$_c$(3900), Y(3940), Z$_c$(4020), Z$_1$(4050), Z$_2$(4250), Y(4140)

\item[$b\bar b$ multiquarks] \hfill \\
Z$_b$(10610), Z$_b$(10650)

\item[other unusual states] \hfill \\
D$_s$(2317), H dibaryon, Y(2175),  Y(4260), Y(4660), Y$_b$(10888), $\pi_1(1600)$, $\pi(1800)$, $f_0(1500)$.
\end{description}
\end{quote}

Although hadronic exotics such as glueballs, hybrids, and multiquark states have been long expected, the understanding of these states is primarily at the level of conjecture. Certainly, if the confirmation of the $Z(4430)$ marks the beginning of the exploration of a new sector of matter, the current phenomenology concerning quark interactions will need to be radically overhauled. A compelling and unified understanding of the new states has not yet emerged, and the gap between theory and experiment remains a major deficiency in our current level of understanding of elementary particle physics.

This gap has its roots in the famously difficult problem of solving QCD in its many-body, strongly interacting, relativistic regime. Current effective field theories are inoperable in the excited spectrum, lattice field theory has difficulties with weakly bound diffuse coupled-channel systems, and extant phenomenological models are insufficiently well constrained to be confidently applied to exotic states. Even the lack of knowledge of relatively simple dynamics, such as interactions in the $K\pi$ system, can affect the analysis of data concerning the new states.

The flood of information initiated by $B$ factories (CLEO, BaBar, Belle), $\tau$-charm facilities (CLEO-c, BESIII), and hadron machines (CDF, D0, LHCb, ATLAS, CMS) is not expected to abate soon. LHCb will continue to deliver new results in heavy quark spectroscopy for at least a decade. At the same time, BESIII at the Beijing Electron Positron Collider will continue its program to collect and analyze $e^+e^-$ data in the energy region of the putative exotic states of charmonium.  Furthermore, the GlueX experiment is due to start taking data in 2015. This experiment, situated at Hall D at JLab, is designed to discover and explore the properties of light hybrid mesons. The COMPASS experiment at CERN has been, and will continue to be, very active in hadron spectroscopy. The PANDA experiment at FAIR is expected to start taking data in 2019; amongst its goals is the exploration of charmonium hybrids and other exotic states.

In view of this situation, a workshop was convened at the Institute of Nuclear Theory, Seattle, with the aim of assessing the status of the field and drawing up a short list of questions that have the potential to move the field forward. This document is the outcome. We stress that this is not meant as a review, for which the reader is directed to Refs.  \cite{Swanson:2006st,Godfrey:2008nc,Brambilla:2010cs,Eidelman:2012vu,Bodwin:2013nua}. Furthermore, the topics contained herein are not meant to be comprehensive, but are offered in the hope that progress will be spurred in various directions.

The next three sections provide specific queries in the areas of lattice field theory, experiment, and theory. The lattice method has been singled out because it has advanced to the stage where modelling issues are minimal, but where results are sufficiently complex that experimental methods must sometimes be invoked to interpret them. Finally, the 
 interface between theory and experiment is addressed in section \ref{thy-int}. Here the emphasis is on smoothing the interaction between theorists and experimental collaborations with the hope of drawing on the strengths of both communities.

\section{Lattice QCD Calculations}

\begin{enumerate}[{1}.1]

\item  \textbf{Compute quantities as a function of light quark mass.} 

A better determination of the contribution of  (virtual)  two heavy-light meson loops in $Q\bar Q$ states below threshold is needed.  Coupled channel phenomenological 
models suggest that for $Q\bar Q$ states near threshold these contributions are significant.  After renormalization of the bare model parameters, one finds modest shifts in the leading nonrelativistic 
mass spectrum.  Spin splittings between ground state heavy-light mesons induce spin-dependent effects in the spectrum and allow hadronic transitions that violate the Heavy Quark Spin Symmetry expectations.  Furthermore, the mass splittings between the $Q\bar u$, $Q\bar d$ and $Q\bar s$ ground state heavy-light mesons allows small isospin breaking and considerable SU(3) breaking effects.
In particular, this may be evident in the large $X(3872) \rightarrow \rho J/\psi$ and $\psi(2S) \rightarrow \eta \psi$ transition rates.

  In lattice QCD calculations,  as the light quark masses are varied down from infinite (quenched approximation)  to the scale of the momentum in the $Q\bar Q$ system
the dominant effect of light quark loops is to modify the running of the QCD coupling ($\alpha_s$).  But as the quark masses are varied below this scale and below $\Lambda_{QCD}$, light quark loops become spatially extended and probe the effects of coupled channels in the hadronic basis of states.  Initial effort in exploring these effects are described in Refs. \cite{sp}.

We suggest that detailed lattice studies of the $Q \bar Q$ mass spectrum as a function of light quark masses ($m_u, m_d, m_s$) for masses in a range between their physical values and
 $\approx 2\times \Lambda_{QCD}$ will give much insight into the effects of coupling to decay channels in a model independent way.   Furthermore,  a calculation of  hadronic transition rates as a function
of light quark masses would be very illuminating.

\item  \textbf{Develop and implement coupled-channel scattering formalism.}

The recent publication \cite{Dudek:2014qha, Wilson:2014cna, Wilson:2015dqa} of the first determinations of \emph{coupled-channel} scattering amplitudes from lattice QCD offers promise that this first-principles approach to QCD might shed light on the exotic behavior being observed in charmonium. For the case of coupled two-body scattering, resonant singularities in the amplitudes can be explored using parameterizations of the $T$-matrix, where the parameters are tuned to describe the finite-volume spectra calculated in lattice QCD. From the pole positions and residues, masses, widths, and branching fractions of resonances can be determined -- the distribution of poles across unphysical Riemann sheets may offer a discriminator for the internal structure of the resonances. 

There has been no application of these coupled-channel techniques to meson systems featuring charm quarks, and only limited studies of elastic scattering, which is a situation in need of remedy. Early targets will be charmed systems near threshold like $DK, D_s \eta$ and $D\pi, D\eta$ as well as exotic isospin and strangeness channels\cite{sp2}. Double charmed channels like $DD$ are also relatively simple. Hidden charm channels are challenging, because while all the tools are in place to deal with the coupled $D\bar{D}, D\bar{D}^*, D^*\bar{D}^*, D_s \bar{D}_s, \ldots$ system, the opening of three-body channels like $\eta_c \pi \pi$ and $J/\psi \pi \pi$ occurs at rather low energies. No complete formalism to relate finite-volume spectra to three-body scattering amplitudes yet exists -- such a formalism will be required to study such systems in detail.

Calculations of meson-baryon scattering are at a less advanced stage than those for meson-meson scattering.
Current stochastic methods for dealing with quark propagation make the calculation of the
$J/\psi\, p$ scattering amplitudes straightforward with a modest increase in cost over meson-meson
amplitudes, even for large volumes\cite{cj}.

\item  \textbf{Investigate static quark interactions.} 

Recent improvements to the set of techniques available for computing light
quark propagation on the lattice should  encourage practitioners to
revisit the problem of computing potentials between static color sources and
their excitation spectra \cite{Perantonis:1990dy,Juge:2002br,Braaten:2014qka,Berwein:2015vca}.
For related work see Refs.  \cite{Guo:2008yz,Szczepaniak:2005xi,Swanson:1998kx}.
These calculations have a long history and the static potential in the $SU(3)$ 
Yang-Mills theory was amongst the first lattice Monte Carlo computations. 
Revisiting the potentials in the presence of light dynamical quarks
\cite{Bali:2005fu,Bicudo:2015vta,Bicudo:2015kna,Brown:2012tm} will give useful
insight into the nature of the $XYZ$ and pentaquark experimental signals. In
particular, the bottom quark sector could be modelled very effectively with
this data while exotic mesons in the charm sector are more sensitive to finite
mass corrections. Phenomenological models of the exotic hadrons based on the
Born-Oppenheimer picture would use these potentials as input.

With a static color and anti-color source, separated along an axis at distance
$R$, the eigenstates of the Hamiltonian are irreducible representations of the
little group of symmetries that preserve this axis.  The energy of these 
states as a function of distance defines the potential, $V(R)$.
The residual symmetry means these potentials are labelled by 
$\Sigma=0^{\pm},1/2,1,3/2,\dots$, where there are two spin-zero potentials 
since a mirror symmetry is also a good symmetry for this 
case. The half-integer spin potentials do not appear in a theory of gluons 
alone but would be present in QCD. 
With two flavors of light quarks, QCD energy eigenstates are classified with 
an extra quantum number, light isospin, and this property would be inherited by 
states built from static sources. There would thus be a new multiplicity of 
spectra with isospin $I=0,1/2,1,\dots$. The isospin 0 and 1 spectra would be 
the relevant ones for studies of hidden charm or bottom tetraquarks and in 
particular, since the $Z^+$ states are charged, the isospin 1 spectrum is of 
particular interest.  This spectrum has not been computed in lattice QCD to 
date. 

For pentaquarks containing hidden charm $c\bar{c}$ or bottom $b\bar{b}$ the 
isospin $1/2$ and $3/2$ potentials are relevant for modelling these states. 
Again, there is no counterpart for this potential in the theory of 
strongly-coupled gluons alone.  Another possible potential that might usefully 
be investigated and which has no counterpart in the pure gauge theory are those 
associated with two color sources, $Q(x)Q(y)$ 
\cite{Bicudo:2015vta,Bicudo:2015kna}. In order to neutralize this 
color charge, at least one light quark field must be included in the creation 
operator. These potentials would help model doubly-charmed or doubly-bottom 
baryons.

\item  \textbf{Compute form factors relevant to exotic states.}

The determination of the elastic and inelastic form factors of the $XYZ$ resonances directly from lattice QCD would have three major impacts. First, it would lead to the theoretical reproduction of experimentally observed production or decay rates in a model-independent way. Second, it will give access to poorly constrained quantities that would elucidate the nature and structure of these exotic states. Examples of such quantities include the radii and electromagnetic moments of tentative molecular states. Third, it will guide future experimental searches of exotics. Although the studies of resonant electromagnetic processes are presently at their early stages, there have been a great deal of theoretical~\cite{Lellouch:2000pv, Briceno:2014uqa, Agadjanov:2014kha, Briceno:2015csa, Briceno:2015tza}
 and numerical~\cite{Dudek:2009kk, Shultz:2015pfa} 
that demonstrate that they are in fact accessible from lattice QCD. This progress resulted in the first calculation of a radiative transition of a hadronic resonance~\cite{Briceno:2015dca}. This calculation was performed in the light sector for $\pi\gamma^\star\to\rho\to\pi\pi$. 

Having determined the $\pi\gamma^\star\to\rho\to\pi\pi$ amplitude for a range of values of the center of mass energy of the final $\pi\pi$ state, the authors of Ref.~\cite{Briceno:2015dca} were able to analytically continue the amplitude onto the $\rho$-pole and determine the $\pi\to\rho$ form factor.

The same technology will be applicable for future calculations in the heavy quark sector.

\item  \textbf{Compute decay constants for exotic states.}

The decay constants of the vector resonances determine their rate of production in $e^+ e^-$ and radiative transitions to lighter states offer a way to produce states of other $J^{PC}$. The rigorously correct way to determine these in lattice QCD is to first determine the scattering amplitudes and their resonant content as described above, and to then introduce an external vector current. By extrapolating the calculated vector-current matrix elements to the resonance poles, off in the complex energy plane, the decay constants and radiative transition rates for resonances can be obtained. 
This procedure closely resembles that for the determination of form factors discussed above. The techniques necessary for implementing this have been previously developed in Refs.~\cite{Meyer:2011um, Feng:2014gba, Briceno:2015csa}. The first calculations of this type has been of $\pi\pi$-electroproduction $\gamma^\star\to\rho\to\pi\pi$~\cite{Feng:2014gba,Bulava:2015qjz}.

A slightly less rigorous approach, which may be acceptable for narrow resonances, is to ignore the hadronic decay of the states by excluding meson-meson-like operators from the basis used to determine the spectrum of states -- a first round of calculations of this simplified type may be justified to aid our phenomenological intuition of the vector spectrum, extending the limited calculations presented in \cite{Dudek:2006ej, Dudek:2006ut, Dudek:2009kk, Becirevic:2014rda, Becirevic:2012dc, Chen:2011kpa}, using the excited state technology presented in \cite{Shultz:2015pfa}.

\item \textbf{In-medium hadron properties.} 

Several model calculations predict that charmonium-nucleus exotic bound states 
should exist~\cite{brodsky_etal,wasson,luke-etal,brodsky-miller,detera,
lee-ko,krein-etal,kazuo-etal,yokota}. Two independent, equally important 
binding mechanisms have been identified: multigluon exchanges in the form of
color van der Waals forces~\cite{brodsky_etal,wasson,luke-etal,brodsky-miller,Brambilla:2015rqa}, 
and $D,D^*$ meson loop contributions to charmonium self-energy with medium-modified 
masses~\cite{lee-ko,krein-etal}. A first, recent lattice calculation~\cite{Beane:2014sda}
confirms model calculation expectations, finding relatively deeply bound states of 
$J/\Psi$ and $\eta_c$ to several light nuclei. Another interesting class of charmed-hadrons 
nuclear bound states are $D$-mesic nuclei~\cite{Tsushima:1998ru,Tolos:2013gta}, which are 
an important source of information on chiral symmetry restoration in-medium~\cite{hatsuda-rmp}.
A lattice calculation of the $D-$meson interaction with nucleons and of $D-$nuclei 
binding energies would be of great importance for constraining models, given the present 
lack of experimental information on the $D-$nucleon interaction.

\end{enumerate}

\section{Experiment}

\begin{enumerate}[{2}.1]

\item  \textbf{Publish upper limits for negative searches.}

Candidates for exotic hadrons have been observed in many channels.  While it is clearly important to find new decay channels for these states, it is also important to limit their decays to other channels when these searches are negative.  This is a reminder to experimentalists to publish upper limits on cross sections and branching fractions in a wide range of channels.

\item  \textbf{Confirm marginal states.} 

A variety of signals have been observed that require confirmation. There is some urgency in achieving this because attempts to understand the data can be seriously misled by the acceptance of spurious signals as hadronic states. Alternatively, many signals are statistically significant but contain unknown systematic errors due to assumptions in modelling (for example using interfering Breit-Wigner amplitudes to obtain asymmetric line shapes). Additional and more varied amplitude analysis is required in these cases.
Amongst states requiring confirmation are $X(3940)$, $Y(4008)$, $Z_1(4050)$, $X(4160)$, $Z_2(4250)$, and $X(4350)$.

\item \textbf{Unravel the excited $\chi_{cJ}$ spectrum.}

The masses of the charmonium $2P$ states are expected to be around
3.8--4.0~GeV/$c^2$~\cite{Barnes:2005pb,Eichten:2002qv}, while
$\chi_{c0}(2P)$ and $\chi_{c2}(2P)$ are well above the $D\bar{D}$
threshold but below $D^*\bar{D^*}$ threshold; they are expected to
be wide. If the mass of $\chi_{c1}(2P)$ is high enough,
$\chi_{c1}(2P)\to D^*\bar{D}+c.c.$ will be its dominant decay
mode. The $\chi_{c2}(2P)$ may decay into $D^*\bar{D}+c.c.$ as well.

So far the $Z(3930)$ observed in $\gamma\gamma\to
D\bar{D}$~\cite{Uehara:2005qd} is regarded as the $\chi_{c2}(2P)$
state, and the $X(3915)$ observed in $\gamma\gamma\to \omega
J/\psi$~\cite{Uehara:2009tx} is supposed to be a $\chi_{c0}(2P)$
candidate, although its mass is a bit too close to
$\chi_{c2}(2P)$, and it was not observed in $\gamma\gamma\to
D\bar{D}$.

Further study on $\chi_{cJ}(2P)$ decaying to $D\bar{D}$ and
$D^*\bar{D}+c.c.$ should be performed to identify $\chi_{c0}(2P)$
and $\chi_{c1}(2P)$ and to confirm the $\chi_{c2}(2P)$.

With more data collected in $e^+e^-$ annihilation at the
$\psi(4040)$ and $\psi(4160)$ peaks, the E1 transitions
$\psi(3S)$ and $\psi(2D)\to \gamma \chi_{cJ}(2P)$ should be
searched for; E1 transitions of $\chi_{cJ}(2P)\to
\gamma\psi(2S)$ are also expected to be large compared with
$\chi_{cJ}(2P)\to \gamma J/\psi$ and $\gamma \psi(1^3D_J)$.

Hadronic transitions $\chi_{cJ}(2P)\to \pi\pi\chi_{cJ}(1P)$ should
be searched for, and the reaction $\chi_{cJ}(2P)\to \omega J/\psi$ may occur 
if the mass difference between $\chi_{cJ}(2P)$ and $J/\psi$ is
large enough. The spin-parity of the $\chi_{c0}(2P)$ candidate,
$X(3915)$, needs to be measured and its production and decay
patterns should be examined carefully to see if it is the
$\chi_{c0}(2P)$.

\item  \textbf{Measure $e^+e^-$ cross sections.}

The region at center-of-mass energies above the open charm
threshold is of great interest due the plethora of vector
charmonium states: the $\psi(3770)$, $\psi(4040)$, $\psi(4160)$,
and $\psi(4415)$ observed in the inclusive hadronic cross section,
and the vector charmonium-like states, the $Y(4008)$, $Y(4260)$,
$Y(4360)$, $Y(4630)$, and $Y(4660)$ observed in exclusive hadronic
modes. These states were discovered in one specific mode and are
not observed in other modes. Searches for these states in all 
possible final states are desired. This suggests high precision
measurements of as many as possible exclusive $e^+e^-$
annihilation modes, including multi-body open charm modes,
hadronic transitions, radiative transitions, and even exclusive
light hadron final states.

Fig.~\ref{xsections} shows an example of measured cross sections
of two-body open charm final states and two- or three-body
hadronic transition modes. Common features of the distributions
are a richness of structures and a lack of precision. With more data from
open charm threshold up to about 5~GeV and improved precision,
better theoretical models will likely be needed to describe the line
shapes of all the final states simultaneously. In this way better
knowledge on the excited $\psi$ and the $Y$ states can be
extracted. This may result in an understanding of the nature of
these states and reveal if any are charmonium hybrids.

\begin{figure*}
\begin{center}
\includegraphics[height=4in]{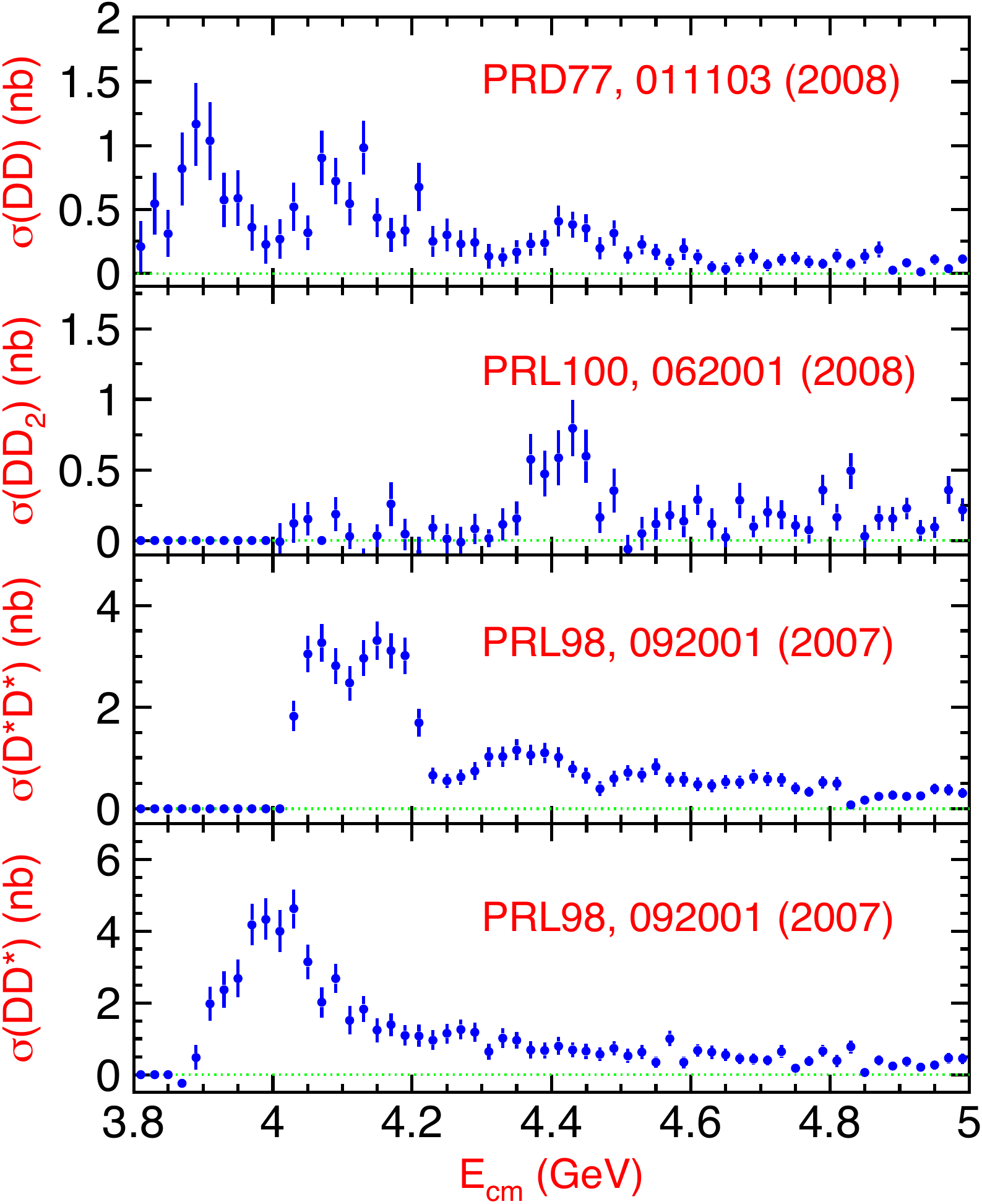}
\includegraphics[height=4in]{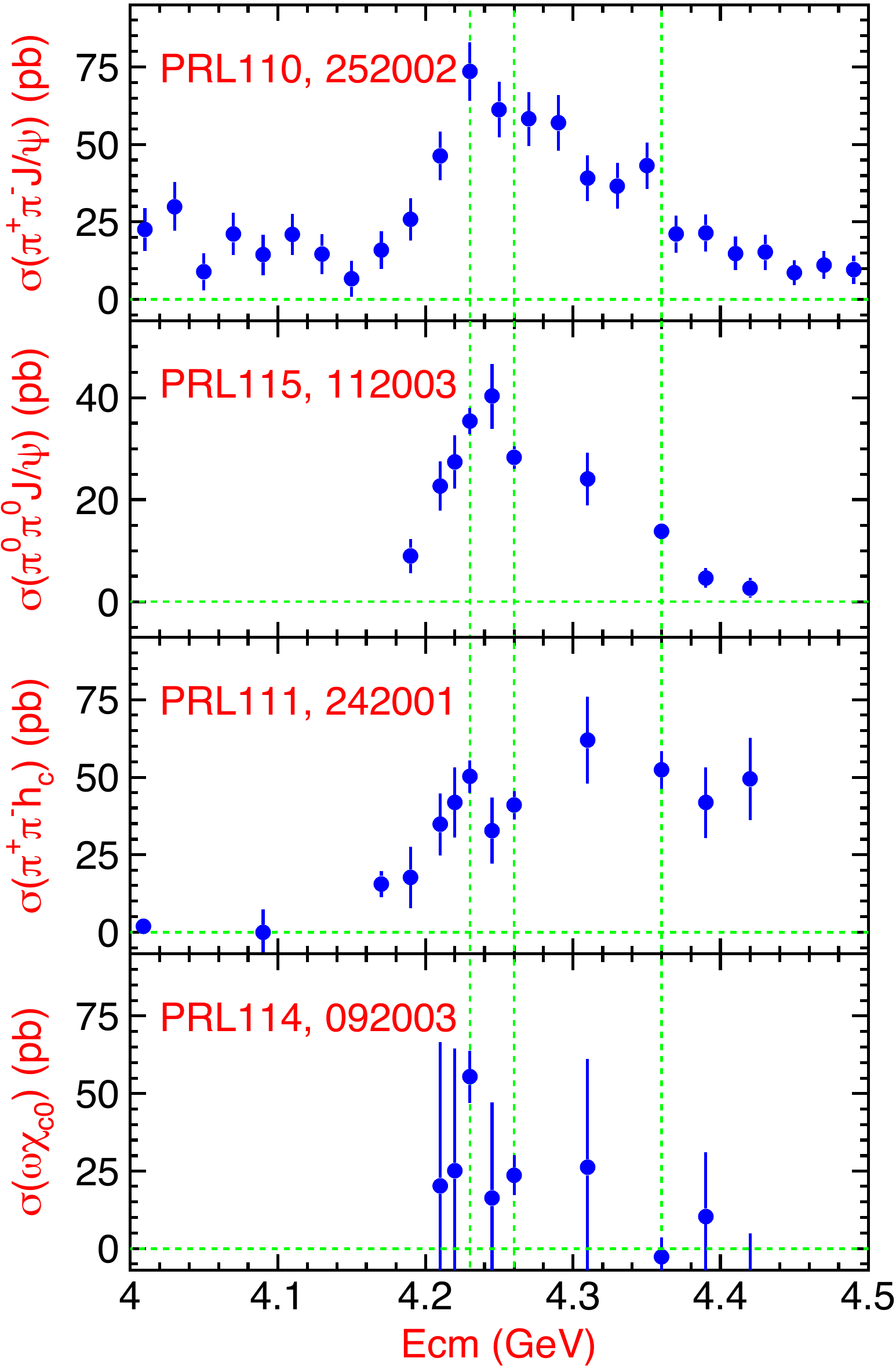}
\caption{The cross sections of $e^+e^-$ annihilation into open
charm final states (left panel, from Belle experiment) and
charmonium final states (right panel, the top is from Belle and
the others from BESIII experiments.) The vertical lines are at
4.23, 4.26, and 4.36~GeV.} \label{xsections}
\end{center}
\end{figure*}

With the existing data samples, BESIII can already improve
precision of the open charm cross sections
significantly~\cite{Yuan:2015kya}, considering the BESIII
experiment will continue run for another 6-8 years, better
measurements at more energy points are expected.
Belle-II~\cite{belle2} will start data taking in 2018 with a data
sample expected to be fifty times larger than Belle's. Thus the precision of
all the measurements with initial state radiation will be
improved.

The cross sections of $e^+e^-$ annihilation into open bottom and
bottomonium final states should also be measured to understand the
excited bottomonium and bottomonium-like states. This can only be
done at the Belle-II experiment~\cite{belle2}.

\item  \textbf{Search for flavor analog exotic states.} 

The majority of recently discovered exotic states are placed firmly in the charmonium spectrum. Flavor-independence of gluon exchange implies that flavor analog states should exist. For example, the $Z_b(10610)$ and $Z_b(10650)$ are evidently hidden bottom partners to the $Z_c(3900)$ and $Z_c(4020)$ multiquark candidate states. It is possible that the $Y(2175)$ is the hidden strange partner of the $Y(4260)$. Finding flavor-analog states will yield valuable information on the dynamics underlying the new states and will probe the robustness of putative models.

The case of a bottomonium analog of the $X(3872)$ is interesting, both because of the novelty of the $X$ and because of differences that may arise. For example, if the $X$ is a weakly bound $D\bar D^*$ system then the $X_b$ would be expected at mass of 10604 MeV. However, some models \cite{Swanson:2003tb} rely on the proximity of the hidden charm $\rho-J/\psi$ and $\omega-J/\psi$ channels to stabilize the $X$. This coincidence is \textit{not} repeated in the case of the $X_b$, where the $\omega-\Upsilon$ threshold lies 370 MeV away. This also implies that the novel isospin-breaking features of the $X$ will not be repeated in the $X_b$ (isospin symmetry breaking is related to the hidden flavor mixing and to the splitting between charged and neutral $D\bar D^*$ channels -- neither of which is repeated in the case of the $X_b$). Finally the proximity of the $\chi_{c1}'$ to the $X(3872)$ is likely to be important. Again, this numerical coincidence is not repeated in the case of the $X_b$, where nearby $\chi_{b1}$ states are at 10255 MeV (1P), 10512 MeV (2P), or 10788 MeV (3P\cite{Godfrey:2015dia}).

\item  \textbf{Search for flavor analogs of the $P_c$.}

The recent evidence for the resonant $P_c^+$ structures in $J/\psi p$ in the $\Lambda_b\to J/\psi p K^-$ decays found by the LHCb experiment
has renewed the interest of the experimental and theoretical communities in pentaquark states.  Further experimental work is critical for clarification of the nature of these structures. 

Most established and candidate exotic hadron states contain hidden heavy flavor, $Q\bar Q$. This is mainly due to experimental constraints for production and detection. However, other sectors of flavor deserve to be investigated.
Let us give two examples.

The isospin partner $(\bar c c u dd)$ and the strangeness partners such as $(\bar c c u d s)$ should be searched for, along with their 
$\bar b b$ analogs. One should not restrict to hidden heavy flavor.
Pentaquark states $(\bar Qq^4)$, where $q^4$ denotes $uuds$, $ddsu$ or $ssdu$ were predicted in 1987 on the basis of a chromomagnetic mechanism very similar to the one leading to speculations about the $H$ dibaryon. This flavored pentaquark has been searched for at Fermilab and HERA. Searches with higher statistics are desirable, especially if more hidden-flavor states such as $(\bar Q q' q^3)$ are found \cite{Gignoux:1987cn,Lipkin:1987sk}.

Exotic mesons with double heavy flavor, $(QQ'\bar q\bar q)$ have been predicted with many methods  such as potential models, QCD sum rules, lattice QCD and the meson-meson molecular picture. It has also been stressed that more effort should be put on double-charm and other doubly-heavy baryons. We thus suggest a search of doubly-heavy hadrons besides $B_c$: double-charm baryons, double-charm mesons and double-charm dibaryons, and in the future, their analogs with charm and beauty or double 
beauty\cite{Ader:1981db,Manohar:1992nd}.

\item  \textbf{Search for quantum number partners of the $Y(4260)$.}

If the $Y(4260)$ is a hybrid state it represents the first example of -- what is expected to be -- a large array of novel hadrons. Specifically, a spin multiplet analogous to those in the conventional spectrum is expected. A lattice computation of the lightest charmonium hybrid multiplets is shown in Fig. \ref{fig-cch}. A clear structure with quantum numbers 
$1^{--}$ and $(0,1,2)^{-+}$ is seen. This multiplet can be conveniently interpreted as arising due to an effective constituent gluon with quantum numbers $(J^{PC})_g = 1^{+-}$ mixing with conventional quark-antiquark degrees of freedom\cite{hsc,Meyer:2015eta}.

\begin{figure*}[h]
\begin{center}
\includegraphics[width=5in]{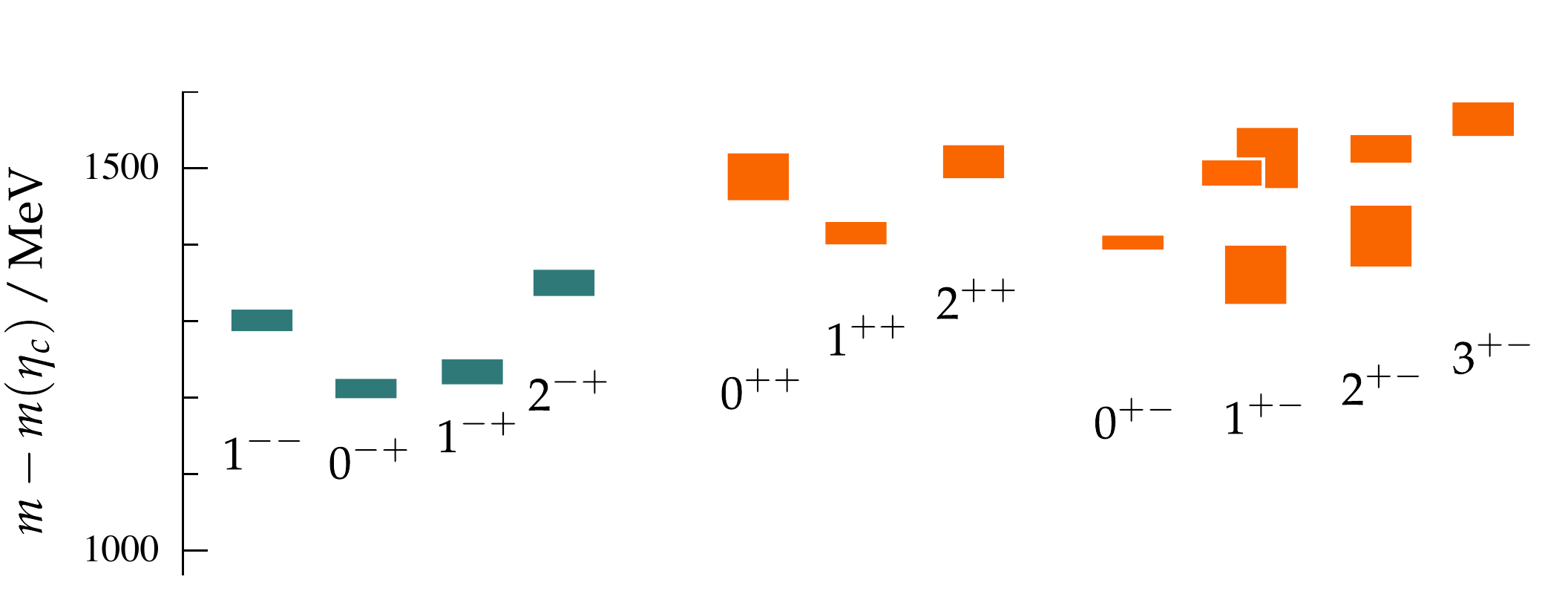}
\caption{The lightest charmonium hybrid multiplets. Based on Ref. \cite{Liu:2012ze}.}
\label{fig-cch}
\end{center}
\end{figure*}

Given this information one can expect the spectrum shown in  Table \ref{tab-cch}.  The increasing luminosity expected at the colliders raises interesting possibilities for detecting these states. For example, the $0^{-+}$ and $1^{-+}$ lie below the $Y(4260)$ and therefore should be accessible in radiative decays in P-wave. The hybrids can then be detected in decay modes such as $\eta\, \chi_{cJ}$ or $f_J\, \eta_c$ (see Table \ref{tab-decays}).

\begin{table}[ht]
\caption{Expected Hybrid Multiplet \cite{Dudek:2009kk}.}
\begin{tabular}{ll}
\hline\hline
$J^{{PC}^{\phantom{x}}}$ & mass (MeV) \\
\hline
$2^{-+}$ & $\sim$ 4320\\
$1^{--}$ & \phantom{$\sim$} 4260 \\
$1^{-+}$ & $\sim$ 4200\\
$0^{-+}$ & $\sim$ 4190\\
\hline\hline
\end{tabular}
\label{tab-cch}
\end{table}

\begin{table}[ht]
\caption{Possible production and decay modes of hybrid charmonium.}
\begin{tabular}{l}
\hline\hline
$Y(4260) \to  \gamma\, 1^{-+} \to \gamma \,\eta\, \chi_{c1}$, $\gamma\, f_1 \,\eta_c$ \\
$Y(4260) \to  \gamma\, 0^{-+} \to \gamma \,\eta\, \chi_{c0}$, $\gamma\, f_0 \,\eta_c$\\
$Y(4360) \to \gamma\, 2^{-+} \to \gamma \,\eta\, \chi_{c2}$, $\gamma\, f_2 \,\eta_c$ \\ 
\hline\hline
\end{tabular}
\label{tab-decays}
\end{table}



\item  \textbf{Pursue properties of the $X(3872)$.}

Although properties of the $X(3872)$ are reasonably well known, additional experimental effort can greatly assist in improving the understanding of this state. For example, the rate for decays to light hadrons, such as $X \to \eta \pi\pi$, can be compared to those for $\chi_{c1}$ states in an effort to determine the expected mixing of the $X$ with the bare $\chi_{c1}(2P)$.

Analog hidden charm states are predicted in some models and can be searched for. For example a $0^{++}$ $D^*\bar D^*$ state is expected at 4019 MeV in pion-exchange models\cite{Swanson:2006st,Tornqvist:1993ng}.

Intriguing analog flavor-exotic states are also expected in QCD. In particular, it has been argued that $QQ\bar q\bar q$ states must exist in the limit where the heavy quark mass goes to infinity\cite{Manohar:1992nd}. The phenomenology of such states is discussed by Tornqvist\cite{Tornqvist:1993ng} and was anticipated long ago\cite{Ader:1981db}.
Specific possibilities include  isoscalar $KK^*$, $DD^*$, $BB^*$ states with $J^P = 1^-$ and $K^*K^*$, $D^*D^*$, $B^*B^*$, etc. Nevertheless, flavor exotic vector-vector bound states are unlikely, except possibly in the doubly charged bottom sector\cite{Tornqvist:1993ng}.

\item  \textbf{Measure additional channels to investigate the $P_c$.}

The interpretation of the LHCb pentaquark signal remains open.
Tightly bound pentaquarks, molecular states, and rescattering effects have been proposed. 
More accurate determination of the quantum numbers of the pentaquark candidates would greatly help their interpretation.
Even before more data is accumulated by the LHCb,
improving parameterizations in the amplitude fits to the existing data may help this end.
For example, models of $\Lambda$ excitations, which dominate the data via decays to $p K^-$, need to be checked for
completeness since the previous experiments may have not discerned all the relevant states, especially at high masses.
Non-resonant terms with slowly varying magnitude and phase can also be significant. 
Alternative approaches to the isobar model may be helpful, like, for example, the recently published  coupled-channel model by Fernandez-Ramirez et al.\cite{Fernandez-Ramirez:2015tfa}. See also Refs. \cite{Burns:2015dwa,Guo:2015umn,Mikhasenko:2015vca}.

It is important to confirm the $P_c$ via other channels. There are already suggestions\cite{Burns:2015dwa} such as $\Lambda_b\to J/\psi p \pi^-$, or $\Xi_b\to J/\psi K^- p$, which are Cabibbo suppressed, or $B\to J/\psi p \bar p$.
The predictions of rescattering models for the $P_c^+$ amplitudes  
can be tested by fitting them directly to the data. 
The presence (or lack thereof) of the same structures in the other channels, 
like $\Lambda_b\to J/\psi p \pi^-$ or $\Lambda_b\to J/\psi p K^0 \pi^-$ is of great importance. 

Rescattering models predict presence of structures related to the $P_c^+$ peaks 
induced by the analyticity in the coupled channels, 
like $\Lambda_b\to \chi_{cJ} p K^-$, $\Lambda_b\to\Sigma^{(*)+}_c\bar{D}^{(*)0}$ and  
$\Lambda_b\to\Lambda^{(*)+}_c\bar{D}^{(*)0}$. 
Ideally, simultaneous coupled channel analysis of the related final states should be performed. 
The investigation of possible structures, which may include depressions rather than peaks, is a good start.
Even total relative rates between the different channels would be interesting.
Negative searches have theoretical implications and should be published. 

Bound-state models for the observed $P_c^+$ states predict other pentaquark states built by the same binding
mechanisms. The same $P_c^+$ states may be observable in the other decay modes too. Thus, every accessible
decay mode of $\Lambda_b$ with $c$ and $\bar c$ quarks among the final state hadrons should be examined, e.g. $\eta_c p K^-$. Other final states can be accessible from $\Xi_b$ to charmonium decays.

Different production mechanisms for the $P_c^+$ states, or their siblings, should be investigated. Examples are prompt production at LHC or photo-production at JLab.

\item  \textbf{Test ideas for meson-nuclear interactions.}

Presently there is a complete lack of experimental information on the low-energy 
interactions of charmed mesons and charmonium with nucleons and nuclei. We look forward 
to several forthcoming experimental programs in this area: the near-threshold 
experiments by the ATHENA collaboration~\cite{ATHENNA} as part of the 12~GeV program 
at Jefferson Lab, the proton-antiproton experiments by the PANDA collaboration at 
FAIR~\cite{PANDA}, and the experiments with 50~GeV high-intensity proton beams at the 
J-PARC complex~\cite{JPARC}. We also envisage opportunities for finding exotic
charmonium-nucleon and charmonium-nucleus bound states with the ongoing heavy-ion
experiments at RHIC and LHC. In particular, we suggest studies on the formation
of such exotic bound states by coalescence in the late-stage evolution of heavy-ion
collisions, as their production yields should be of comparable magnitude to those
of anti-nuclei and anti-hypernuclei recently observed at RHIC~\cite{Abelev:2010rv} 
and LHC~\cite{Adam:2015vda}.

\item  \textbf{Improve meson classification scheme.}

There is a wide range in signal robustness in the spectrum of new states. Because this can lead to confusion amongst theorists who are attempting phenomenological descriptions of these new states, we recommend that a star system for mesons be implemented by the Particle Data Group for use in the \textsl{Review of Particle Properties}. Furthermore, the current exotic particle naming scheme is somewhat confused and is applied inconsistently; we therefore recommend that a consistent and flexible nomenclature be implemented.

\item  \textbf{Search for $p\overline{p}$ in decays at LHC for PANDA.} 

Heavy-flavor physics will benefit from experiments with medium-energy antiproton beams. In the past, a precursor signal of the $h_c$ was seen at the CERN ISR, and many properties of the $\chi_{c,J}$ and other charmonium states were obtained from the $\bar p p$ experiment at the Fermilab accumulator. 

To assist future experiments, it is desirable to get information on the coupling of heavy hadronic systems to proton-antiproton pairs by detecting $\bar p$ production at $B$-factories and at the LHC. This is already under way, and this should be accompanied by more theoretical studies. For instance, it remains rather mysterious that $\eta_c(1S)$ decays more often to $p\bar p$ than suggested by simple perturbative QCD, while $\eta_c(2S)$ is more weakly coupled to that channel.

\end{enumerate}

\section{Theory and Phenomenology}

\begin{enumerate}[{3}.1]

\item \textbf{Study exclusive $e^+e^-$ cross sections using better coupled-channel formalism.}

The identification of possible new resonances implied by the $XYZ$ phenomena requires studies of analytical amplitudes that describe the relevant production and decay characteristics\cite{Szczepaniak:2015Dalitz,Szczepaniak:2015eza}.
  For example, in the case of the $Z_c(3900)$ that is observed in the $\pi^\pm J/\psi$ spectrum in decays of the $Y(4260)$ to $\pi^+\pi^- J/\psi$,  the relevant direct channels involve the nearby  open-charm, $D\bar D^* + {c.c}$ states and need to be included in a coupled channel formalism.  The open charm resonances in the production channel, e.g., the $D_1(2420)$ can produce rapid variations of the direct channel partial waves near the $Z_c$ signal and need to be taken into account in production.  The singularity structure of partial wave amplitudes is constrained by unitarity, therefore 
  a comprehensive analysis requires implementation of unitarity constraints in all relevant channels. 
   This requires simultaneous studies of quasi two-to-two scattering amplitudes of  open flavor and 
    heavy quarkonia, {e.g.} $D\bar D^* \to J/\psi \pi$, and eventually a study of three-to-three scattering, {i.e.} $D\bar D \pi \to D \bar D \pi$ amplitudes.

\item \textbf{Develop tests for the dynamical diquark picture.} 

In an alternate proposal for the structure of the heavy quarkonium-like
exotics, both for the tetraquarks~\cite{Brodsky:2014xia} and
pentaquarks~\cite{Lebed:2015tna}, the states are composed of compact
diquark-antidiquark (-antitriquark) pairs rapidly separating and
hence ultimately achieving large ($\approx 1$~fm or greater)
separation before decay.  This picture has features in common with the
diquark models previously
mentioned~\cite{Maiani:2004vq,Maiani:2014aja}, but differs in that the
states are extended, dynamical rather than compact, static objects,
and therefore does not necessarily admit a Hamiltonian description.
Nevertheless, in the limit of small separation, the two pictures
should coincide.  The first priority in this case is therefore the
development of a formalism in which the spectrum can reliably be
predicted.  A first attempt in the pentaquark
sector~\cite{Zhu:2015bba}, still using a Hamiltonian formalism, gives
a natural explanation for a broad ${\frac 3 2}^-$ lying just below a
narrow ${\frac 5 2}^+$, consistent with the LHCb
findings~\cite{Aaij:2015tga}, but also predicts a large number of
undiscovered states.  A lattice calculation of the potential
corresponding to a well-separated static diquark-antidiquark pair may
provide valuable information on the possible spectrum.  Since the
exotics are so prominent in the $c\bar c$ and $b\bar b$ channels, some
hints of the same mechanism with $s\bar s$ (hidden-strangeness
pentaquarks) should appear in processes such as $\Lambda_c \to \phi
\pi^0 p$~\cite{Lebed:2015dca} or $\phi$
photoproduction~\cite{Lebed:2015fpa}.  A primary benefit of the
dynamical picture is its natural explanation of strong overlaps with
spatially larger states, so a precision measurement of the ratio
$Z(4475) \to \psi(2S) \pi$ vs.\ $J/\psi$ and to other states will be
illuminating.  The dynamical and compact diquark models share an
expected enhancement of $Z_c \to \eta_c \rho$ compared to the
corresponding rate in molecular models~\cite{Esposito:2014hsa}. The
extended structure of the state may also offer interesting
opportunities for the production of unusual final-state particle
correlations.  The multiparticle nature of states produced via, say,
electroproduction or $p \bar p$ annihilation (at JLab or PANDA, respectively) can be probed by means of constituent counting
rules~\cite{Brodsky:2015wza}, and can help to distinguish whether
compact multiquark components are produced.

\item  \textbf{Develop experimental tests for tetraquarks.} 

Compact tetraquark configurations, in which all four quarks
participate in strong mutual interactions, can be distinguished from
the hadron molecular picture or threshold effects through a variety of
experiments.  The most well developed tetraquark models are of the
diquark-antidiquark class~\cite{Maiani:2004vq,Maiani:2014aja}, and
rely on Hamiltonians with spin-spin interactions between the quark
pairs.  A comparison of the expected spectra in this tetraquark model
versus hadronic molecular models (and also
hadrocharmonium)~\cite{Cleven:2015era} indicates that many more states
should arise if tetraquarks are the dominant exotic component; for
example, the $X(3872)$ should have isotriplet charged partners of the
same $G$ parity [and opposite that of the $Z(3900)$ and $Z(4020)$].
Due to the proximity of thresholds, such states might exist only as
very broad yet-undiscovered resonances.  Large prompt production cross
sections at colliders~\cite{Chatrchyan:2013cld} argue against
$X(3872)$ being a $D \bar D^*$ molecule forming through coalescence;
indeed, an extrapolation~\cite{Esposito:2015fsa} of data from ALICE
shows that production of loosely bound hadronic molecules such as $d$
and ${}^3$He at high $p_\perp$ will be quite suppressed, unlike
current indications for $X(3872)$, an effect that can be decisively
checked in future ALICE and LHCb experiments.  The molecular and
diquark pictures also differ radically in the ratios of their
branching fractions of $Z_c \to \eta_c \rho$ vs.\ $J/\psi \pi$ or $h_c
\pi$~\cite{Esposito:2014hsa,Li:2014pfa}, the former being dozens of times
less frequent in molecular models.  Loosely bound molecules also must
obey well-known universal relations (independent of the potential)
between binding energy and width, and precision measurements of the
resonance widths and constituent masses can help determine whether
these constraints are satisfied~\cite{Polosa:2015tra}.

\item \textbf{Develop techniques for $5q$ and $6q$ systems} 

Potential models provide some guidance for QCD calculations. Two-body calculations are obvious once an explicit potential is given. Three-body and four-body computational methods new yield accurate spectra, although they require more delicate tools. The  case of five-body and six-body systems are still debated. For instance, with similar Ansatze for the interaction, the $H=(uuddss)$ can be found to be either stable or unbound. We suggest to publish a set of benchmark calculations to remove the ambiguities.

\item \textbf{Pursue the Born-Oppenheimer method (adiabatic surface mixing).} 

The presence of heavy charm or bottom quarks in the new putative tetraquark mesons 
suggests that they may be successfully studied using the
\textit{Born-Oppenheimer expansion}.
This approach was introduced by
Born and Oppenheimer in 1920  \cite{Born-Oppenheimer}
to understand the binding of atoms into molecules
by exploiting the large ratio between the mass of an atomic nucleus 
and an electron, which implies that
the time scale for the motion of electrons is orders of 
magnitude faster than that for the motion of the nuclei.
The energies of stationary states of the electrons 
in the presence of fixed nuclei 
can be calculated as functions of the separation of the nuclei.
The resulting functions are called {\it Born-Oppenheimer potentials}. 
In the Born-Oppenheimer approximation,
these functions are used as potential energies in the 
Schr\"odinger equation for the nuclei,
under the assumption that the electrons respond very rapidly
to the motion of the nuclei.
The Born-Oppenheimer {\it expansion} involves taking
the large mass ratio into account more systematically
by incorporating non-adiabatic couplings between different
stationary states of the electrons.
This results in coupled-channel Schr\"odinger equations 
that systematically improves the description of a molecule.

The Born-Oppenheimer expansion was  
applied to mesons containing a heavy quark ($Q$) 
and antiquark ($\bar Q$) in 1999\cite{Juge:1999ie}, exploiting the fact
that, since the mass of the heavy quark 
is much larger than the typical energies of the gluons and light quarks,  
the time scale for the evolution of the gluon and light-quark fields is much faster 
than that for the motion of the $Q$ and $\bar Q$. 
In Ref.~\cite{Juge:1999ie}, lattice QCD was used to calculate the 
Born-Oppenheimer potentials defined by the energies of the gluons 
in the presence of fixed $Q$ and $\bar Q$ as functions 
of the $Q \bar Q$ separation.
These energies were then used as the potential energies in the
Schr\"odinger equation for the $Q$ and $\bar Q$.  
The bound states in the Born-Oppenheimer potentials
were interpreted as meson resonances.  
These bound-state energies were compared 
with corresponding meson masses computed directly using lattice QCD
and agreement in the level spacings to within 10\% was found, 
strongly supporting the validity of the Born-Oppenheimer 
expansion for such systems. 

The approach used in Ref.~\cite{Juge:1999ie} should be extended
to apply to the $XYZ$ mesons and to
include nonadiabatic effects that can be incorporated
through coupled-channel Schr\"odinger equations.
The Born-Oppenheimer potentials for heavy tetraquark mesons
and the nonadiabatic couplings between the potentials 
could be calculated using lattice QCD.  The heavy quark and
antiquark would be treated as static, and the energies of the gluons
and light quarks could then be computed as a function of the
separation between the quark and the antiquark.
The resulting coupled-channel Schr\"odinger equations 
could then be solved to determine the energies and widths of resonances,
which can be compared with the observed $XYZ$ mesons, and possibly to 
predict new tetraquark mesons.

\item  \textbf{Revisit conventional meson models.}

While the successes of the constituent quark model are well-known in the heavy quark sector, the efficacy of the model is not expected to survive higher in the spectrum, where gluonic and coupled channel effects become important. Of course, it should be possible to extend constituent models to include these additional degrees of freedom, but experimental and theoretical guidance will be required. 

Even the simple problem of assessing the accuracy of the constituent quark model above threshold has difficulties. For example, there are eight charmonium states below $D\bar D$ threshold that are all well-described by models. Alternatively, the situation above threshold is considerably more confused; of the approximately twenty claimed states, most of them are not understood, and even well-known states such as the $\psi(3770)$ lie 50 MeV below the prediction of the Godfrey-Isgur model. In the bottom sector the 14 states that lie below $B\bar B$ are well-described. In this case there are only six states above threshold, but, again, the experimental and theoretical situation is confused.

Since the new experimental data lie firmly in the continuum region, it is very likely that more sophisticated versions of the quark model that respect unitarity must be developed. Of course, this has been known in the community for many decades, and much work has been done\cite{Eichten:ag,Eichten:1978tg,Eichten:1979ms,Torn,Ono:1983rd,Heikkila:1983wd,Tornqvist:1984fy,Zenczykowski:1985uh,Geiger:yc,Geiger:ab,Geiger:va,Morel:2002vk,van Beveren:2003af,van Beveren:2004bz,van Beveren:2004ve,Amsler:2004ps,Kalashnikova:2005ui,Hanhart:2007yq,Kalashnikova:2007qz,Eichten:2004uh,Barnes:2007xu,Pennington:2007xr,Rupp:2015taa}.  There are daunting issues to be overcome, including determining the form of the nonperturbative gluonic transition operator and evaluating the (divergent) sum over infinitely many virtual channels\cite{Swanson:2005rc}. Nevertheless it is difficult to imagine progress being made without a successful outcome to this effort. Alternative approaches exist of course: lattice gauge theory is rapidly making progress in working in the coupled channel regime, and one hopes that effective field theory approaches will be developed that can accommodate the extra scales present.

\item  \textbf{Develop the Dyson-Schwinger Formalism.}

The Dyson-Schwinger equations (DSEs) of QCD, together with various many-body equations 
for bound states (Bethe-Salpeter equations for the two-body problem, Faddeev 
and Faddeev-Yakubovsky equations for the three- and four-body problem) have the 
potential to reveal the connections between the physics in different sectors of QCD.
The equations encode the running of QCD Green's functions (for example, the quark mass 
function) and therefore connect the perturbative current quark region with 
the non-perturbative constituent quark domain. Furthermore, they connect the
heavy quark regime, where NRQCD or potential models are applicable, with the 
light quark sector, where the concept of a potential is not very well defined. 

The explanatory power of the DSE framework with respect to exotic hadrons 
is still in its early exploration stage. So far, light scalar mesons have been treated as tetraquarks in an approach that takes into account two-body correlations within the bound state 
equation for two quarks and two antiquarks\cite{Heupel:2012ua,Eichmann:2015cra}. The resulting 
Bethe-Salpeter amplitude for scalar tetraquarks is dominated by pseudoscalar 
meson-meson correlations. For the lightest state, the $f_0(500)$, this explains 
its large decay width into two pions, whereas the $a_0$ and $f_0$ 
are dominated by their $K\bar{K}$ components. In general, it turns out that 
all two-body correlations inside the tetraquark (i.e., (anti-)diquarks or mesons) 
contribute to the wave function and it is a question of the internal dynamics which
is the dominant cluster. For the light scalar mesons this is the `meson molecule'
configuration, but other results are in principle possible for other quantum numbers 
and different quark flavors and masses.

Whether this mechanism has the potential to shed some light
on the question of the internal structure of the tetraquarks among the $XYZ$-states, 
in particular their (anti-)diquark, molecular or hadrocharmonium nature, needs to explored. To this end, non-scalar quantum numbers need to be studied and the framework
needs to be extended toward heavy-light systems. Furthermore, more quantitative 
precision is needed to confirm the prediction of an all-charm tetraquark in the
5.0 - 6.5 GeV mass region \cite{Heupel:2012ua,Eichmann:2015cra}.

Complementary ongoing projects within the DSE framework concern the glueball spectrum 
\cite{Meyers:2012ka,Sanchis-Alepuz:2015hma} and the question whether states with 
exotic quantum numbers can be accounted for with relativistic quark-antiquark 
systems (in contrast to the non-relativistic quark model) \cite{Hilger:2015hka}.

\item  \textbf{The status of Large $N_c$ Considerations.} 

One striking thing about modern exotics--- the $XYZ$ states---is that they all involve the physics of heavy quarks.    This raises an interesting issue: are heavy quarks necessary for the formation of exotics or do exotics exist for light quarks systems.  The experimental data on this is murky.  The large $N_c$ limit may provide a bit of insight.   The subtle point is that the large $N_c$ and heavy quark limits may not commute so that generic large $N_c$ arguments based on scaling arguments really apply for light quark systems.  The  standard version of the large $N_c$ limit with quarks in the fundamental representation of  SU($N_c$) can be shown not to have narrow tetraquarks at large $N_c$\cite{Cohen:2014tga}, apparently supporting the proposition that the heavy quarks are  necessary for the existence of tetraquark states.  However, there is a variant of the large $N_c$ limit where quarks are in the two-index anti-symmetric representation of SU($N_c$) in which it can be shown that states with exotic tetraquark quantum numbers must exist as narrow resonances ({i.e.} states whose widths go to zero as $N_c$ goes to infinity) regardless of the mass of the quarks\cite{Cohen:2014via}.  Minimally this shows that QCD-like gauge theories are not excluded from  having tetraquark states even if the quarks are light.  

\end{enumerate}

\section{Theory-Experiment Collaboration}
\label{thy-int}

The following are a few suggestions that could help facilitate collaboration between theory and experiment.

\begin{enumerate}[{4}.1]

\item \textbf{Improve parameterizations of the data.}

One of the challenges in many of the experimental studies of the $XYZ$ states is to develop correct parameterizations of the data.  For example, amplitude analyses often find a need to introduce non-resonant terms. At present, very little theoretical guidance is provided except for the amplitude formulations based on the K-matrix approach. However, the latter is not always practical.

To improve this situation, we have two recommendations.

First, we encourage that, when appropriate and beneficial, experimentalists and theorists directly work together on the analysis of data.  This could be accommodated by theorists becoming co-authors on specific experimental papers they substantially contributed to, or joint submission of experimental and theoretical papers cross-referencing each other. The experiments are encouraged to formalize procedures making such collaboration possible, and theorists are encouraged to approach the experiments when they think they might directly aid specific data analysis topics.  Further progress could be made by more persistent forms of collaboration, including direct involvement of theorists in the data analysis process within the established procedures of the experimental collaborations.

Second, we encourage theorists, when possible, to publish complete functional forms (amplitudes, etc.) that could be used in the fitting of data.  One example of this is in the parameterization of rescattering amplitudes. The current theoretical calculations are dependent only on the center of mass energy ~\cite{Liu:2015Pc, Mikhasenko:2015vca, Szczepaniak:2015Dalitz}, whereas amplitudes used in fitting require a flexible parameterization involving the angular information of the decay. If theorists develop more complete rescattering amplitudes, experimentalists could use them in analyses. This would most likely involve some collaborative effort in understanding both how the experimental analysis is performed and what are the theoretical requirements for such amplitudes.  Another example is in resonance parameterizations: it would be useful for experimentalists if a number of alternate resonance parameterizations were available that could be used in systematic studies.

\begin{figure*}
\begin{center}
\includegraphics[width=3in]{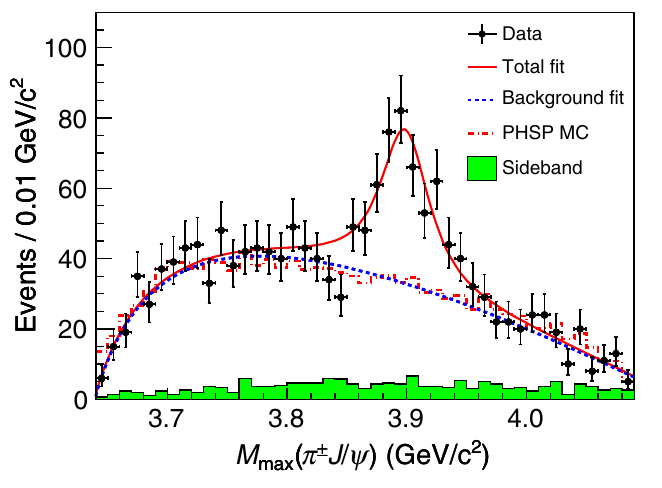}
\includegraphics[width=3in]{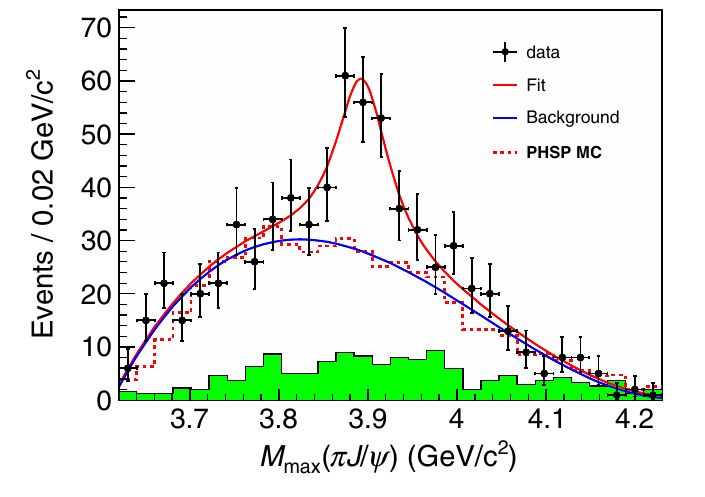}
\caption{The observation of the $Z_c(3900)$ from BESIII~\cite{Ablikim:2013mio} (left) and from Belle~\cite{Liu:2013dau} (right).  The different shapes at low $M(\pi^{\pm}J/\psi)$ mass are partly due to differences in experimental detection efficiencies.} \label{zcshapes}
\end{center}
\end{figure*}

\item \textbf{Make experimental results more accessible for subsequent interpretation.}

The analysis of data from many modern experiments often necessitates complying with internal rules designed to provide collaborative controls over the quality of statistical methods used and the proper evaluation of systematic uncertainties. Therefore it is unrealistic to expect that all data will be made available for analyses outside of this collaborative setting.  A correct analysis of data would benefit from the types of closer interaction between experiment and theory discussed in the previous point.

A different issue is how published data (for example, Dalitz plots) should be subsequently interpreted.  It often occurs that experimental results are made public in a manner that does not easily allow for subsequent interpretation.

One example is the discovery of the $Z_c(3900)$ decaying to $\pi^\pm J/\psi$ in the process $e^+e^-\to\pi^+\pi^- J/\psi$~\cite{Ablikim:2013mio,Liu:2013dau}.  The data presented in the discovery papers include several effects that are difficult to take into account when performing theoretical fits.  First, the BESIII data~\cite{Ablikim:2013mio} was taken at $\sqrt{s}=4.26$~GeV, while the Belle data~\cite{Liu:2013dau} includes a range of energies around the $Y(4260)$ peak.  This makes the Belle data, in particular, hard to subsequently fit, since any changes (beyond the size of available phase space) in the $\pi^\pm J/\psi$ mass spectrum as a function of $\pi^+\pi^-J/\psi$ mass are unknown. Second, the two experiments have different experimental efficiencies over the Dalitz plot due to differing detectors and kinematics.  These effects are not quantified in the publications.  The importance of these two effects can be seen when comparing the $M(\pi^\pm J/\psi)$ plots from BESIII and Belle, which differ substantially, especially in the low-mass region~(Fig.~\ref{zcshapes}).  It is therefore not clear how one could correctly analyze the published data with various new parameterizations to test, for example, differences between cusp and resonant models of the $Z_c(3900)$.

When deemed appropriate, experiments are therefore encouraged to make efficiency-corrected data available for external analyses. Or, when possible or desired, experiments could make published plots publicly available along with efficiency curves and instructions for how to use the plots for subsequent analysis.  This could be provided as supplemental information to a publication.  This may be easy for simple three-body final states (like $\pi\pi J/\psi$, where the Dalitz plot could be provided), but impractical for more complicated final states.

Another suggestion, especially when amplitude analyses have been performed, is to publish a complete parameterization of the data, including both the formulas and the numerical values for each fit parameter.

Making data public in these ways could help facilitate ongoing efforts to test and build models.  It could also permit combined fits of data from different experiments or different channels, thus helping a more global picture to emerge.

\item  \textbf{Preview upcoming analysis results.}

It may be useful, in some circumstances, for experimental collaborations to provide a list of upcoming experimental results.  This might be a fruitful way to elicit new theoretical predictions or ideas.  For example, if the community knows a certain measurement is being performed, there is then a chance to make predictions prior to the publication of experimental results.  It also permits setting priorities in theoretical computations and enhances the possibility of arranging collaborative effort. Hosting such a list on a common platform may prove useful and should be explored.

\item \textbf{Create a list of publications in $XYZ$ physics.}

It would be helpful to have a centralized running bibliography of publications relevant for $XYZ$ physics.

\end{enumerate}

\acknowledgments

We are grateful to David Kaplan and the INT for financial and logistics support. The authors thank the following agencies for financial support:
the U.S. Department of Energy (Cohen);
the Institute of Modern Physics and Chinese Academy of Sciences under contract Y104160YQ0 and agreement No.~2015-BH-02 (Coito);
the U.S. Department of Energy, for grant DE-AC05-06OR23177, under which Jefferson Science Associates, LLC, manages and operates Jefferson Laboratory and  DE-SC0006765, Early Career award (Dudek);
Fermilab, operated by the Fermi Research Alliance under contract number DEAC02-07CH11359 with the U.S. Department of Energy (Eichten);
BMBF, under contract No. 06GI7121, and the DAAD under contract No. 56889822 
and by the Helmholtz International Center for FAIR within the LOEWE program of the 
State of Hesse (Fischer);
the German Research Foundation DFG under contract number Collaborative Research
Centre CRC-1044 (Gradl);
the Conselho Nacional de Desenvolvimento
Cient\'{\i}fico e Tecnol\'ogico - CNPq, Grant No. 305894/2009-9 and
Fun\-da\-\c{c}\~ao de Amparo \`a Pesquisa do Estado de S\~ao Paulo - FAPESP,
Grant No. 2013/01907-0 (Krein);
U.S. National Science Foundation, under grants PHY-1068286 and PHY-1403891 (Lebed);
the Brazilian National Council for Scientific and Technological Development under grant  CNPq/CAPES-208188/2014-2 (Machado);
U.S. Department of Energy under grant DE-FG02-05ER41374 (Mitchell);
U.S. National Science Foundation under grant PHY-1306805 (Morningstar);
U.S. Department of Energy, supported by Jefferson Science Associates, LLC under contract No. DE-AC05-06OR23177 (Pennington);
the National Natural Science Foundation of China (NSFC) under contract No. 11575017, 
the Ministry of Science and Technology of China under Contract No. 2015CB856701 (Shen);
U.S. Department of Energy, under grant  DE-FG02-05ER41374 (Shepherd);
U.S. National Science Foundation under grant PHY-1507572 (Skwarnicki);
U.S. Department of Energy, under contract DE-AC05-06OR23177 and grant DE-FG0287ER40365 (Szczepaniak);
the National Natural Science Foundation of China (NSFC) under contract numbers 11235011 and 11475187 (Yuan).

 \end{document}